# SCFNet: A Transferable EEG IIIC Classification Network

Weijin Xu

East China Jiao Tong University

2022028081100031@ecjtu.edu.cn

## Abastract

Epilepsy and epileptiform discharges are common harmful brain activities, and electroencephalogram (EEG) signals are widely used to monitor the onset status of patients. However, due to the lack of unified EEG signal acquisition standards, there are many obstacles in practical applications, especially the difficulty in transferring and using models trained on different numbers of channels. To address this issue, we proposes a neural network architecture with a single-channel feature extraction (Singal Channel Feature) model backend fusion (SCFNet). The feature extractor of the model is an RCNN network with single-channel input, which does not depend on other channels, thereby enabling easier migration to data with different numbers of channels.

Experimental results show that on the IIIC-Seizure dataset, the accuracy of EEG-SCFNet has improved by 4% compared to the baseline model and also increased by 1.3% compared to the original RCNN neural network model. Even with only fine-tuning the classification head, its performance can still maintain a level comparable to the baseline. In addition, in terms of cross-dataset transfer, EEG-SCFNet can still maintain certain performance even if the channel leads are different.

## 1  Introduction

Epilepsy is a chronic brain disorder caused by abnormal discharges of neurons in the brain; epileptiform discharges (IED) refer to epileptic-like discharges such as spikes or sharp waves detected on an electroencephalogram (EEG), but without concurrent clinical seizures, this pattern of brain discharge is also referred to in the medical field as the ictal-interictal continuum (IIC, seizure period-interictal period continuum) or ictal-interictal-injury-continuum (IIIC, seizure period-interictal period-injury continuum). Epileptic seizures and epileptic brain activity patterns can damage the brain, and in severe cases, may even lead to death.

EEG detection is an effective method for monitoring the onset status of such diseases and is widely used in clinical medicine. In past clinical detection and classification tasks, manual methods were commonly used; however, manual diagnostic methods require specially trained medical experts, are time-consuming and labor-intensive, and rely on the personal experience of experts. To improve efficiency, various automatic EEG-based epilepsy recognition methods have been developed. Early methods often used manually extracted features for recognition, some based on morphology, others on feature engineering, but due to the complexity of the clinical environment, such as significant signal interference and missing signal interruptions, the reliability of manual features is poor. Recently, with the development of artificial intelligence, deep learning has gradually been applied to epilepsy detection.

In recent years, deep learning models have achieved remarkable success in the automatic diagnosis of biological signals[1], including sleep stage classification[1-3], emotional analysis[4], motion image analysis[5], and EEG-based epilepsy seizure classification[6-9]. Some studies use one-dimensional convolutional neural networks (CNN) on raw signals[10-12], while others preprocess data using short-time Fourier transform (STFT) and apply two-dimensional CNN models on the generated spectrograms[13-15]. Researchers also segment signals and use CNN segment encoders with downstream sequences, such as transformers or recurrent neural networks (RNN), to capture temporal dynamics[16-19]. Other methods include ensemble learning and feature fusion using multiple encoders[20].

In the field of automatic epilepsy detection, literature [21] divides filtered EEG signals into smaller segments and calculates the histogram of 1D-LBP for these segments, finally using a nearest neighbor classifier for classification, achieving an accuracy rate of 98.33% on public datasets. Literature[22] tests the performance of two universal time series networks, InceptionTime and Minirocket, on TUEV and private datasets, with AUC, AUPRC, and F1 scores of 0.99, 0.99, and 0.97, respectively, on the TUEV dataset. Literature[23] collects a six-class IIIC dataset and uses a one-dimensional Densnet as the classification network for raw signal detection and classification, with experimental results showing it outperforms trained experts on ROC and PRC metrics. Literature[24] uses a two-dimensional CNN network for the time-frequency map of EEG detection and classification, with an average accuracy rate of 89.63% in 10-fold cross-validation on the CHB-MIT dataset.

However, these models mainly focus on fixed-format sequential signal samples for specific tasks, greatly limiting the network's migration capabilities. Especially in EEG acquisition, there are significant differences in acquisition standards, equipment, and data formats, mainly reflected in the different positions of leads, sampling rates, and numerical units. Currently, deep learning models for EEG signals are usually designed for specific datasets and clinical environments, limiting their widespread application.

This paper proposes a single-channel feature extraction backend fusion network architecture based on the single-channel detection method in sleep detection. This model architecture divides multi-channel data into independent channel data and uses a single-channel input RCNN network to extract features from each channel, effectively solving the problem of model coupling with the number of channels in EEG detection. The main contributions of this study are as follows:

The proposed SCFNet model architecture, by transforming the model structure while maintaining the original feature extraction model structure, achieves decoupling from the number of EEG channels while maintaining comparable accuracy. Pre-trained models can achieve transfer learning across different channels and datasets by only fine-tuning the classification head.

The EEG-SCFNet and IIT-SCFNet models, transformed from EEGNet and IITNet[25], can reduce the number of training epochs from 6-8 to 2-4 during training, effectively improving training efficiency and convergence speed.

## 2 SCFNet

### 2.1 Model Overview

The inspiration for the SCFNet model comes from two aspects. Firstly, IITNet[25] uses a single channel to analyze and classify sleep activities using EEG, and some other sleep papers also use a single channel to classify brain activities. Secondly, SPaRCNet [26] uses a method of swapping channel positions to transform and expand the dataset. From these two points, it can be inferred that the EEG waveforms of each channel have independent and identically distributed characteristics.

The model architecture of single-channel feature extraction followed by fusion classification mainly consists of a single-channel feature fusion module and a classifier module. The single-channel feature fusion module uses an RCNN module architecture of InceptionLayer-ResNet1d-BiLSTM, and the classifier uses a two-layer MLP. The feature extraction module can also use other architectures, such as pure CNN1D, Transformer, or other machine learning models for single-channel extraction.

The network architecture proposed in this paper is significantly different from traditional end-to-end methods in terms of the organization design of the feature extraction layer and the classification layer. Traditional methods usually treat multi-channel time series data segments as a whole sample for processing, while this paper adopts an innovative strategy: splitting each data segment's multi-channel into multiple independent single-channel data segments. As shown in Figure 1, traditional end-to-end learning uses the feature extraction process to treat signals from different channels as a whole, with multiple channel signals being fused during compression. Although this method can learn the relationships between multiple channels well, the number of channels and the order of channels are fixed. The proposed method splits each channel data segment into independent channel data, treating each channel's data segment as a sample and performing feature extraction on each channel. This method gives up learning the relationships between channels, but each channel is independent, and channels can be added or removed as needed without retraining the feature extractor.

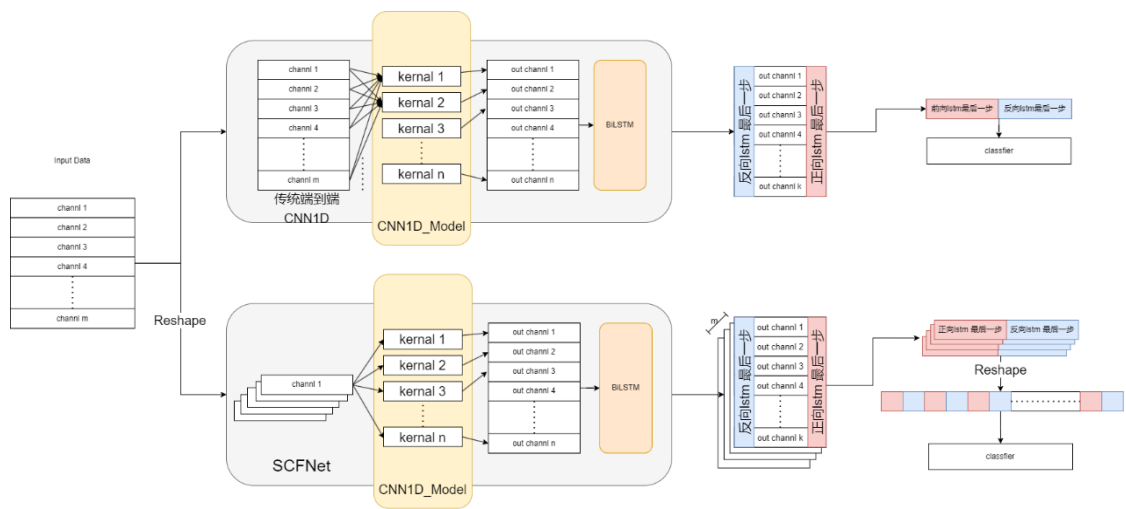

Figure 1 Traditional end-to-end architecture versus SCFNet architecture

As shown in Figure 2, the single-channel feature extraction module (Single Channel Feature Module, SFC Module) has no essential difference in architecture from traditional end-to-end networks, only requiring the number of input channels to be set to 1. The single-channel feature extraction module used in this paper adopts the RCNN model, which is composed of two main modules: CNN convolutional layers and RNN recurrent neural network layers. The convolutional layer is constructed by stacking one Inception1D layer and nine ResNet1D layers. The design of the Inception1D layer is inspired by the field of image classification and contains four parallel convolutional networks with kernel sizes of 3, 5, 7, and 9, respectively. From a temporal perspective, smaller convolutional kernels have a smaller field of view and focus on capturing shorter temporal features, while larger convolutional kernels focus on longer temporal features; from a frequency perspective, small convolutional kernels are more sensitive to high-frequency features, and large convolutional kernels are more concerned with low-frequency features. This design cleverly balances time and frequency precision, achieving comprehensive feature extraction of temporal data. The ResNet1D module draws on the ResNet architecture from image processing, effectively suppressing gradient explosion and vanishing problems by introducing residual blocks, allowing the model to be stacked deeper. At the same time, due to the introduction of residuals, the model is easier to reduce training errors when adding new layers, thereby improving the accuracy and efficiency of feature extraction. Finally, the RNN part of the RCNN model uses a Bi-LSTM network, which is a typical recurrent neural network that can better handle sequential data, further enhancing the model's performance. Overall, the network consists of Inception, Resnet1d, and BiLSTM, which will be referred to as EEGNet in the following text.

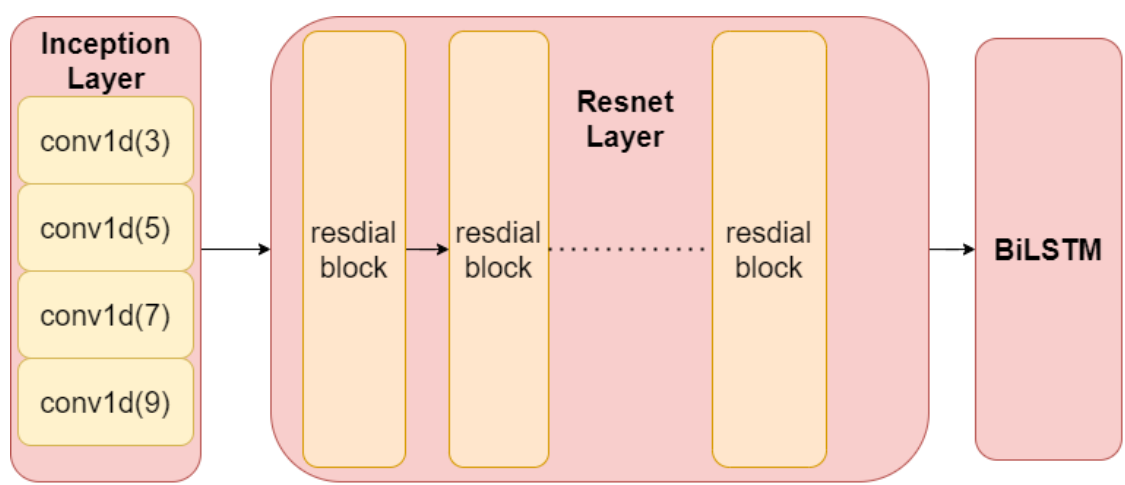

Figure 2 Structure of EEGNet model

## 2.2 Model Modification Method

While the basic model possesses the capability to classify and learn from data, there exists a tight coupling relationship between the number of input channels in its feature extraction module and the number of data channels. This means that once the number of input channels changes, the previously trained model based on that data will not be able to function normally or be transferred. However, the model structure proposed in this paper, which separates channels and fuses features at the backend, effectively solves this problem. As described in the overview of the model, this structure allows for independent feature extraction from each channel, without relying on information from other channels. This characteristic greatly enhances the reusability of the feature extraction module, enabling the model to easily adapt to data with different numbers of channels, thereby facilitating convenient transfer. Moreover, since the number of channels in each sample causes the feature extraction module to be trained correspondingly multiple times, this actually speeds up the training of the feature module and significantly reduces training time.

The proposed model structure not only solves the limitations of traditional models when the number of input channels changes but also improves the efficiency and reusability of feature extraction through channel separation and backend feature fusion, providing greater flexibility and convenience for model transfer and application.

Transforming a basic model into an SCFNet model is very simple; it only requires changing the number of input channels of the traditional end-to-end model to 1 and modifying the forward computation process. As shown in Figure 3, this is the forward computation process of an SCFNet. In the specific training process, first, the same feature extraction module is used to extract features from each single-channel data segment. This means that if the original data has n channels, then the feature extraction module will be called n times in each training, processing each channel separately. Compared with the traditional end-to-end method, this step significantly increases the usage frequency of the feature extraction module. After completing feature extraction, we concatenate the features extracted from each channel to form a complete feature vector, which is then input into the subsequent classifier for classification prediction.

From the perspective of code implementation, the key to this method lies in reshaping the data. In the initial stage, the shape of a single data sample is channels×step, but before feature extraction, we need to reshape it to channels×1×step to meet the input requirements of the feature extraction module. During batch training, the data shape is further adjusted to (batch_size×channels)×1×step to process multiple channels of multiple data samples simultaneously. The training process of the transformed model is shown in Figure 3, where m is the number of input data channels, and the hidden layer dimension of BiLSTM is the same as the output channel number of the CNN network, both are k.

The prediction formula of the traditional end-to-end model can be written as Formula (1), while the improved formula can be written as Formula(2). The original feature extractor relied on multiple channels for modeling, while the transformed one only relies on a single channel, with all the knowledge learning between channels transferred to the classifier model.

$$Predict = F_{class}[f(x_{ch1}, x_{ch2}, \ldots, x_{chn})] \quad (1)$$

$$Precdict = F_{class}[f(x_{ch1}), f(x_{ch2}), \ldots, f(x_{chn})] \quad (2)$$

In summary, the network architecture proposed in this paper achieves effective classification of time series data by splitting multi-channel data segments, performing feature extraction separately, and concatenating feature vectors. This method not only improves the flexibility of feature extraction but also provides richer feature information for the classifier. The main advantages of SCFNet are fast convergence and convenient migration.

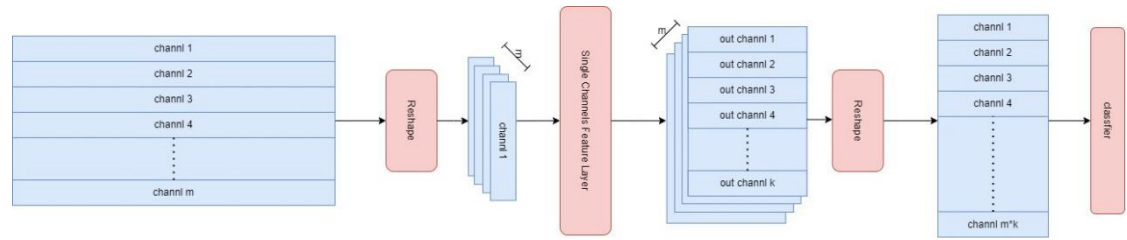

Figure 3 Overall architecture of EEG-SCFNet

## 3 Experiments and Results

To verify the transferability of the modified model, this study conducted transfer learning experiments on different channel count data and cross-dataset transfer learning experiments on the IIIC-Seizures dataset and the CHB-MIT dataset. The IIIC-Seizures dataset is a multi-classification task dataset for studying epilepsy and epileptic brain activities, while the CHB-MIT dataset is a binary classification task dataset for studying epilepsy. The binary classification task of CHB-MIT is a subtask of the multi-classification of IIIC-Seizures, so it is feasible to transfer the model trained on IIIC-Seizures to CHB-MIT.

### 3.1 Dataset Introduction

The IIIC-Seizure dataset is a collection of epilepsy and epileptiform data gathered by Harvard Medical School, classified using the IIIC classification standard. It comprises six categories: Seizures, LPD (Lateralized Periodic Discharge, focal periodic discharges), GPD (Generalized Periodic Discharge, generalized periodic discharges), LRDA (Lateralized Rhythmic Delta Activity, focal rhythmic delta activity), GRDA (Generalized Rhythmic Delta Activity, generalized rhythmic delta activity), and Other (difficult to distinguish). The IIIC-Seizure dataset includes 50,697 labeled EEG samples and 6,095 EEG samples from 2,711 patients, annotated by doctors and experts from 18 institutions. The dataset has a sampling rate of 200 Hz and includes 17 channels: Fp1-F7, F7-T3, T3-T5, T5-O1, Fp1-F3, F3-C3, C3-P3, P3-O1, Fp2-F4, F4-C4, C4-P4, P4-O2, Fp2-F8, F8-T4, T4-T6, T6-O2, and EKG (electrocardiogram signal). The dataset has two types of labels: hard labels, which use independent one-hot encoding to classify the six categories, and soft labels, which use expert voting numbers to mark the data.

The CHB-MIT dataset is a collection of EEG recordings from pediatric subjects with intractable epilepsy. Subjects underwent monitoring for several days after withdrawing from antiepileptic drugs to describe their seizure characteristics and assess their suitability for surgical intervention. The dataset includes 969 hours of scalp EEG recordings and 173 seizures. The dataset contains various types of seizures (clonic, atonic, tonic) and the diversity of patients (male, female, aged 10-22) and different types of seizures make it suitable for evaluating the performance of automatic seizure detection methods in real-world environments. The dataset has a sampling rate of 256 Hz and includes common channels FP1-F7, F7-T7, T7-P7, P7-O1, FP1-F3, F3-C3, C3-P3, P3-O1, FP2-F4, F4-C4, C4-P4, P4-O2, FP2-F8, F8-T8, T8-P8-0, P8-O2, and other channels, totaling 23 channels. The dataset is a binary classification problem with labels for seizures and non-seizures.

## 3.2    Dataset Analysis and Data Preprocessing

Due to the soft labels in the IIIC-Seizure dataset being based on the number of expert votes, the number of data reviewers varies for different samples. Table 1 shows the distribution of various types of samples under different numbers of experts (the number of experts indicates cases greater than that value). It can be seen that the data distribution changes when the number of experts is greater than 3 and greater than 5. According to the dataset paper, a high-quality dataset subset consists of cases with more than 10 experts, and this segmentation method is adopted in this experiment. The IIIC-Seizure dataset also has an extremely imbalanced epilepsy category. However, oversampling for sample balance training did not yield good results, so no sample balancing operation was performed.

Table 1 IIIC-Seizure Data distribution for different expert numbers in the dataset

| 专家人数 | GPD | GRDA | LPD | LRDA | Other | Seizure |
|---|---|---|---|---|---|---|
| 0 | 16702 | 18861 | 14856 | 16640 | 18808 | 20933 |
| 1 | 15851 | 18524 | 13824 | 16445 | 16973 | 20823 |
| 2 | 15625 | 18340 | 13341 | 16378 | 15878 | 20562 |
| 3 | 11854 | 5609 | 9802 | 6999 | 11464 | 2529 |
| 4 | 11087 | 5540 | 8542 | 6633 | 10904 | 2100 |
| 5 | 10451 | 5402 | 7645 | 6289 | 10407 | 638 |
| 6 | 10077 | 5383 | 7440 | 6230 | 10220 | 599 |
| 7 | 10077 | 5383 | 7440 | 6230 | 10220 | 596 |
| 8 | 10077 | 5383 | 7440 | 6230 | 10220 | 596 |
| 9 | 10077 | 5383 | 7440 | 6230 | 10220 | 596 |
| 10 | 10011 | 5099 | 7199 | 5800 | 10112 | 579 |
| 11 | 9836 | 4525 | 6601 | 5059 | 9611 | 566 |
| 12 | 9314 | 3777 | 5841 | 4030 | 8384 | 496 |
| 13 | 7108 | 2811 | 4461 | 3184 | 6324 | 429 |
| 14 | 5975 | 2531 | 3166 | 2875 | 5521 | 362 |
| 15 | 2198 | 1214 | 1518 | 1239 | 3371 | 225 |
| 16 | 681 | 419 | 696 | 129 | 2585 | 64 |
| 17 | 620 | 295 | 415 | 61 | 1691 | 47 |

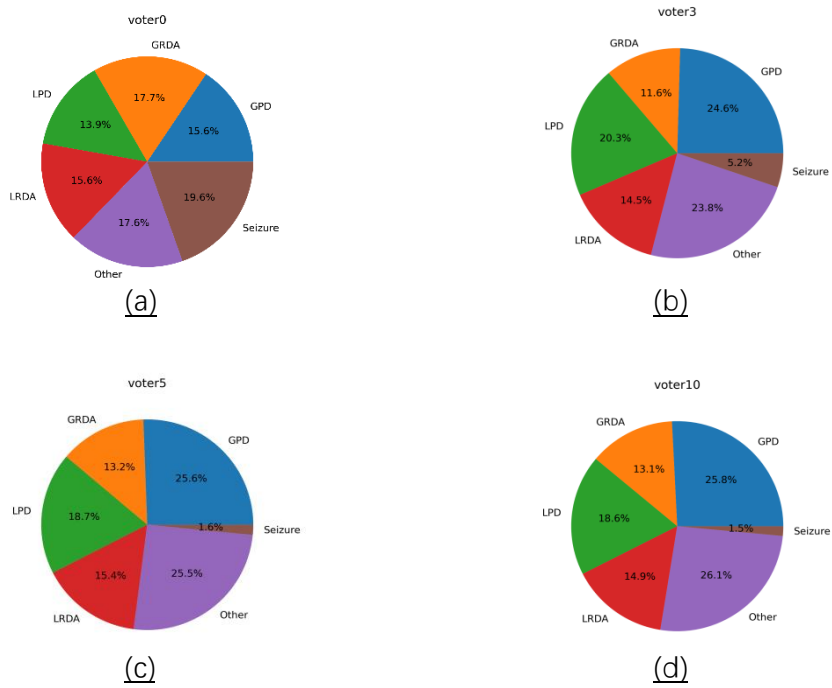

Figure 4 IIIC-Seizure Data distribution in different vote counts of the dataset

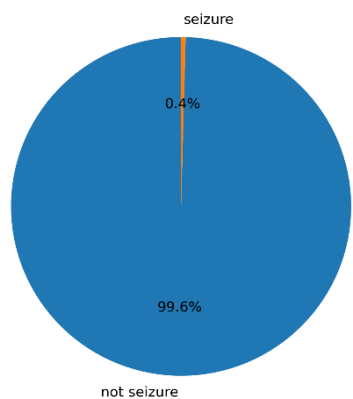

Figure 5 Proportion of epileptic to non-epileptic data in the CHB-MIT dataset

Due to the different sampling frequencies of the IIIC-Seizure and CHB-MIT datasets, which are 200Hz and 256Hz respectively, resampling techniques are used to downsample the CHB-MIT dataset to 200Hz. The CHB-MIT dataset contains 23 channels, but not all data channels are the same. Sixteen common channels are selected, and after processing, 31 files with different channels are removed from the CHB-MIT dataset, leaving 655 files. Subsequently, the data from both datasets is segmented into samples with a duration of 50 seconds each. As shown in Figure 5, there is a scarcity of epilepsy data, accounting for only 0.4%. When training the network using the CHB-MIT dataset, oversampling methods are used for sample balance training.

Data Augmentation

Data augmentation can generate virtual samples, producing more diverse data and thus enhancing the network's generalization capability. Article [27] studied the contribution of different data augmentation methods to contrastive learning of ECG signals. The research found that the combination of random resized crop (RRC) and random signal truncation (Time Out) methods is most effective. Based on this study, we use these two methods for data augmentation of EEG data.

As shown in Figure 6(a) Random Resized Crop: Randomly adjust the size and crop a random continuous segment of the signal, then resize it back to its original size. We uniformly sample the crop parameter p from the range (l, m), where (l, m) are the transformation parameters. In our experiments, we use (l, m) = (0.8, 1.0), meaning we crop the signal to a part between 80%-100%.

As shown in Figure 6(b) Random Signal Truncation: Set a random continuous segment of the signal to zero. It accepts a range (t_l, t_u) as parameters, from which the timeout parameter t is uniformly sampled. This parameter describes how much of the signal will be set to zero. In our experiments, we use (t_l, t_u) = [0, 2000], so we set up to 2000 points of the signal to zero.

In addition, we applied the channel random swap augmentation method, which randomly swaps the positions of two channels within a hemisphere and randomly swaps the order of channels between the two hemispheres. For example, if channels 1-8 are the left hemisphere and channels 9-16 are the right hemisphere, then the order of channels 1-8 in the left brain is randomly changed with probability, the order of channels 9-16 in the right brain is randomly changed with probability, and the order of the left and right brains is randomly swapped with probability.

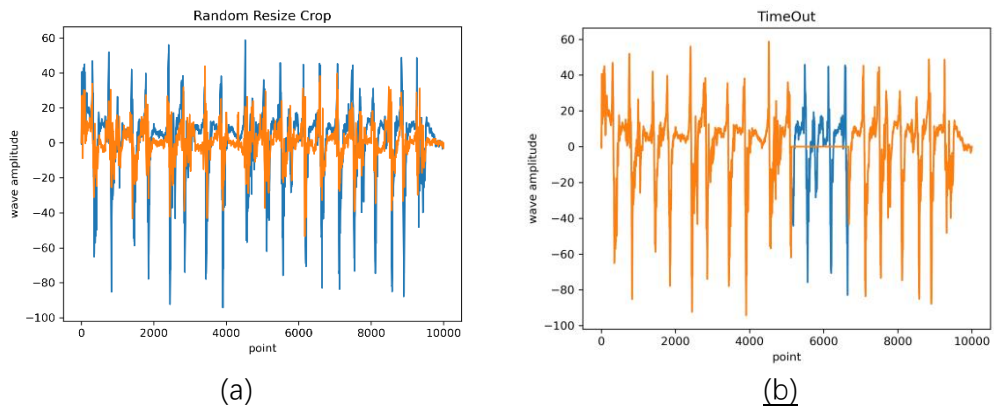

Figure 6 RRC and TO data enhancement methods

## 3.3 Experiment Design

This experiment uses the K-fold cross-validation method to train and evaluate the model. K-fold cross-validation is a widely applied model evaluation technique in machine learning and data mining, aimed at reducing the variance of model evaluation through multiple trainings and tests, thus improving the efficiency and accuracy of model selection. K-fold cross-validation randomly divides the dataset into K equally sized subsets (called "folds"), then sequentially uses each fold as the test set, and the remaining K-1 folds as the training set for model training and evaluation. This process is repeated K times, each time selecting a different fold as the test set, and finally, the average of the K evaluation results is taken as the overall performance indicator of the model. In this K-fold experiment, the data is divided by patient group to prevent data leakage (or data pollution) and overfitting, thereby improving the accuracy and reliability of model evaluation.

The IIIC-Seizure dataset has two types of labels: hard labels and soft labels. Hard labels use one-hot encoding to classify six categories, while soft labels use the number of votes to mark the data. Since the IIIC-Seizure provides a test dataset on Kaggle, which uses KLDivLoss as the evaluation criterion, targeting soft labels, this experiment uses soft labels for supervised training of the network, using KLDivLoss as the loss function. The training data uses a subset with more than 10 votes for 5-fold cross-validation training and evaluation.

The CHB-MIT dataset is a binary classification problem, so CrossEntropyLoss is used as the loss function. Similarly, 5-fold cross-validation is used for training and evaluation.

To verify the model's performance and two types of transferability (channel

transformation transferability and dataset transferability), this experiment includes several parts:

1. Train models using 16-channel and 8-channel pairs of IIIC-Seizure data and compare their performance.
2. Use the model trained on IIIC-Seizure for transfer learning on CHB-MIT and compare model performance.

Since the binary classification task of the CHB-MIT dataset is a subset of the IIIC-Seizure task, CHB-MIT is mainly used to verify the test set of dataset transfer performance.

## 3.4 Metric I

## 3.5 ntroduction

KLDivLoss, also known as KL divergence loss, is a loss function used in deep learning to measure the difference between two probability distributions. It evaluates the model's performance by calculating the KL divergence between the input probability distribution and the target probability distribution. Mathematically, it is defined as follows, where p(xi)$p(xi)$ and q(xi)$q(xi)$ are the probability values of distributions P and Q at the i$i$-th data point, and N is the number of samples:

$$D_{KL}(P||Q) = \sum_{i=1}^{N} p(x_i) \cdot log\left(\frac{p(x_i)}{q(x_i)}\right) \quad (3)$$

CrossEntropyLoss, or cross-entropy loss function, is suitable for classification problems and is used to measure the difference between two probability distributions. The formula is as follows, where i$i$ is the number of classes:

$$L = -\sum_{i} y_i \log(\hat{y}_i) + (1 - y_i)\log(1 - \hat{y}_i) \quad (4)$$

Micro TPR (True Positive Rate): Micro true positive rate is an indicator for evaluating the performance of a classification model, especially important when dealing with imbalanced datasets. For convenience in subsequent charts, "acc" is used to refer to this indicator. The calculation formula is as follows, where C$C$ is the number of classes:

$$Micro\ TPR = \frac{\sum_{i=1}^{C} TP_i}{\sum_{i=1}^{C} TP_i + \sum_{i=1}^{C} FN_i} \quad (5)$$

ROC AUC (Receiver Operating Characteristic Area Under the Curve) is an indicator used to evaluate the performance of a binary classification model. It measures the model's classification ability by plotting the ROC curve and calculating the area under it. Macro

AUC and Micro AUC are two important indicators for evaluating the performance of a multi-class classification model. They measure the model's classification ability by calculating the average AUC of each class and the overall AUC of all samples, respectively. The calculation formulas are as follows, where nclasses$n_{classes}$ is the number of classes, and npairs,total$n_{pairs,total}$ represents the total number of all possible class pairs. In multi-class problems, if there are $k$ classes, then there will be $k(k-1)/2$ different class pairs; $r_i$ and $r_{i-1}$ represent the true positive rates (TPR) of the class pairs; $f_i$ and $f_{i-1}$ represent the false positive rates (FPR) of the class pairs:

$$Macro\ AUC = \frac{1}{n_{classes}} \sum_{i=1}^{n_{classes}} AUC_i \quad (6)$$

$$Micro\ AUC = AUC_{micro} = \frac{1}{n_{pairs,total}} \sum_{i=1}^{n_{pairs,total}} (r_i - r_{i-1}) \times (f_i - f_{i-1}) \quad (7)$$

Experimental The hardware platform for this experiment consists of Nvidia 4080 and Intel 12700, while the software platform encompasses Windows 11, Docker, Ubuntu 22.04.5 LTS, Conda, Python 3.10, and PyTorch. The experiment is categorized into the training results on the IIIC-Seizure and CHB-MIT datasets, respectively, to verify the cross-channel transfer learning ability and cross-dataset transfer learning ability. The training results on the IIIC-Seizure dataset, including those on 16-channel and 8-channel data, exhibit the cross-channel transfer learning ability of EEG-SCFNet. The training results on the CHB-MIT dataset illustrate the cross-dataset transfer learning ability of EEG-SCFNet from IIIC-Seizure to CHB-MIT. The training results on the IIIC-Seizure dataset

## 3.6   The training results on the IIIC-Seizure dataset

In this experiment, the model Densnet employed in the IIIC-Seizure dataset paper was taken as the benchmark model. IITNet and the EEGNet proposed herein were respectively utilized as the prototype models and transformed into IIT-SCFNet and EEG-SCFNet. The experimental results are presented in Table 2. EEGNET_M (EEG-SCFNet) is the modified model based on the fundamental model, EEGNET is the basic model, IITNET_M (IIT-SCFNet) is the model modified based on IITNET, and Densnet is the model utilized in the IIIC-Seizure dataset paper. In the 8-channel experiment section, EEGNET_M_F (EEG-SCFNet-Finetuning) and IITNET_M_F (IIT-SCFNet-Finetuning) are fine-tuning models that use the 16-channel pre-trained channel model. After freezing the parameters except for the classifier, only the classifier is trained. chn represents the number of channels utilized, and time indicates the data duration. acc is the micro-true positive rate of k-fold cross-validation, and private_loss and public_loss are the scores on Kaggle using KLDivLOss, where smaller values are preferred. Epoch refers to the total number of training rounds for the 5-fold training. The model provided in the original paper has a private_loss of 0.73 and a public_loss of 0.87 on the Kaggle test set. Since the official model employed the entire dataset as the training set, it was not evaluated using the training approach in this study.

Table 2 Comparison of model training results on the IIIC-Seizure dataset

| 模型名称 | chn | time/s | acc/% | private_loss | public_loss | Epochs | Macro_AUC/% | Micro_AUC/% |
|---|---|---|---|---|---|---|---|---|
| EEGNET_M | 16 | 50 | 67.34 | 0.41 | 0.35 | 16 | 89.45 | 91.22 |
| EEGNET | 16 | 50 | 66.09 | 0.44 | 0.37 | 38 | 87.35 | 90.1 |
| IITNET_M | 16 | 50 | 65.53 | 0.4 | 0.35 | 33 | 87.95 | 88.95 |
| IITNET | 16 | 50 | 63.36 | 0.5 | 0.43 | 42 | 84.2 | 88.4 |
| Densnet | 16 | 50 | 63.61 | 0.51 | 0.45 | 31 | 83.48 | 86.89 |
| EEGNET_M_F | 8 | 10 | 65.88 | 0.45 | 0.37 | 11 | 87.9 | 89.61 |
| EEGNET | 8 | 10 | 61.79 | 0.48 | 0.38 | 24 | 85.69 | 88.82 |
| IITNET_M_F | 8 | 10 | 63.59 | 0.46 | 0.39 | 30 | 86.53 | 88.91 |
| IITNET | 8 | 10 | 61.02 | 0.53 | 0.45 | 50 | 84.2 | 88.4 |
| Densnet | 8 | 10 | 56.89 | 0.53 | 0.45 | 52 | 81.24 | 86.12 |

From the perspective of accuracy, it can be seen from the table that the proposed modified model can improve model performance. EEG-SCFNet outperforms EEGNET, and IITNET-SCFNet outperforms IITNET, and is 2%-3% higher than the baseline model Densnet. When migrating from 16-channel data to 8-channel data for transfer learning, EEG-SCFNet can still maintain performance with only the classifier trained. From the perspective of convergence speed, it can be seen that EEG-SCFNet and EEG-SCFNet-Finetuning use the least number of training rounds on 16-channel and 8-channel data, respectively, which are 16 and 11 times, far lower than other models. Compared with the original models, IITNET-SCFNet-Finetuning and EEGNET-SCFNet-Finetuning also reduce the number of training rounds.

From the above analysis, it can be seen that the single-channel feature extraction model structure constructed based on the basic EEGNet model in this paper can effectively improve performance and can be effectively migrated to data with different channels.

### 3.6.1  EEG-SCFNet performance

The above demonstrates the transfer learning capability of SFCNet across different channels. This section specifically showcases the performance of EEG-SCFNet on the

IIIC-Seizures dataset. As shown in Figure 6(A), the ROC curve of EEG-SFCNet using 16-channel 50s data segments, and Figure 6(B) is the confusion matrix. It can be seen that the proposed EEG-SFCNet performs very well in three classes: GPD, LPD, and Other, but performs poorly in LRDA and GRDA, and very poorly in the Seizure class. The specific TPR values are shown in Table 3.

From Figure 6(C), which shows the Macro ROC curve, and Figure 6(D), which shows the Micro ROC curve, it can be seen that the proposed EEG-SFCNet achieves the optimal values for both Macro ROC and Micro ROC parameters.

Table 3 Comparison of training results between EEG-SCFNet model and benchmark model

| 模型 | Acc | seizure | lpd | gpd | lrda | grda | other |
| --- | --- | --- | --- | --- | --- | --- | --- |
| EEGNet_M | 0.67 | 0.15 | 0.64 | 0.82 | 0.40 | 0.48 | 0.84 |
| EEGNet | 0.66 | 0.17 | 0.60 | 0.74 | 0.50 | 0.51 | 0.82 |
| Densnet | 0.64 | 0.03 | 0.51 | 0.63 | 0.49 | 0.11 | 0.86 |

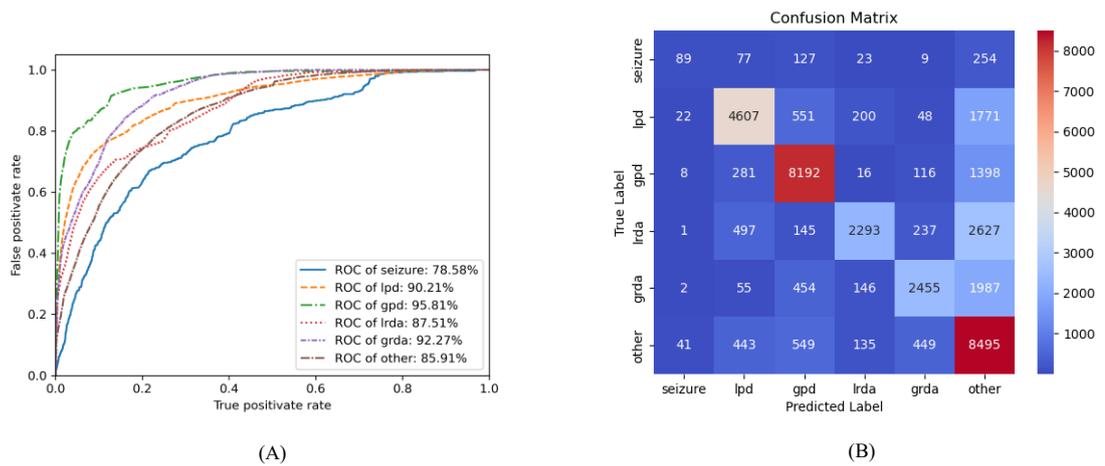

(A)　　　　　　　　　　　　(B)

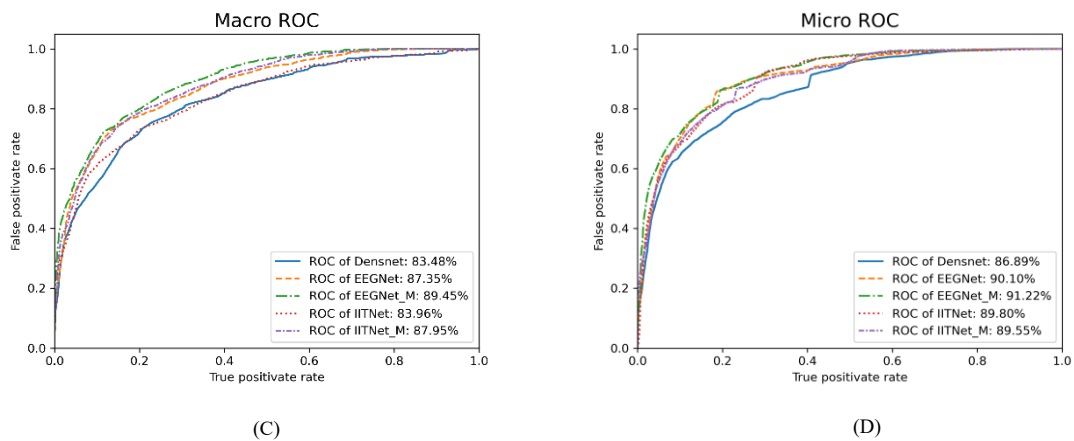

(C)　　　　　　　　　　　　(D)

Figure 7 EEG-SCFNet performance

Figure 8 shows the t-SNE dimensionality reduction visualization of EEG-SCFNet, where each point represents a sample marked on the two-dimensional plane after dimensionality reduction. In Figure 9, red indicates the sample marks corresponding to the respective category, and blue indicates the sample marks of non-corresponding categories. 0 to 5 correspond to Seizures, LPD, GPD, LRDA, GRDA, and Other, respectively.

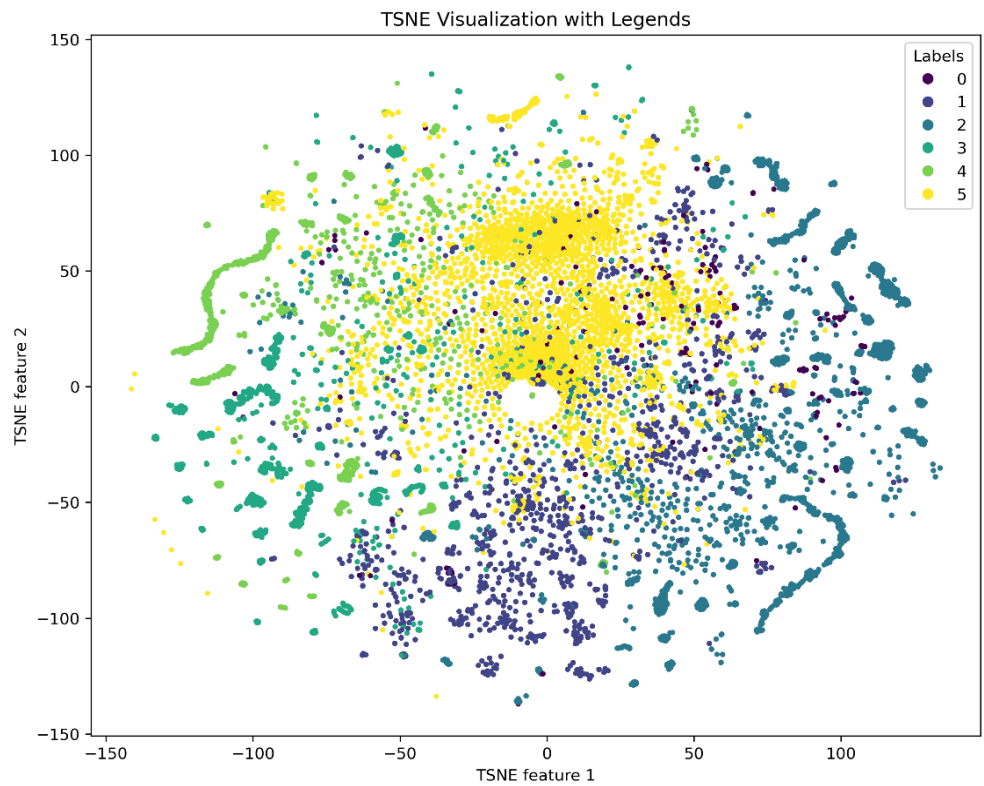

Figure 8 t-sne dimensional reduction visualization of EEG-SCFNet model

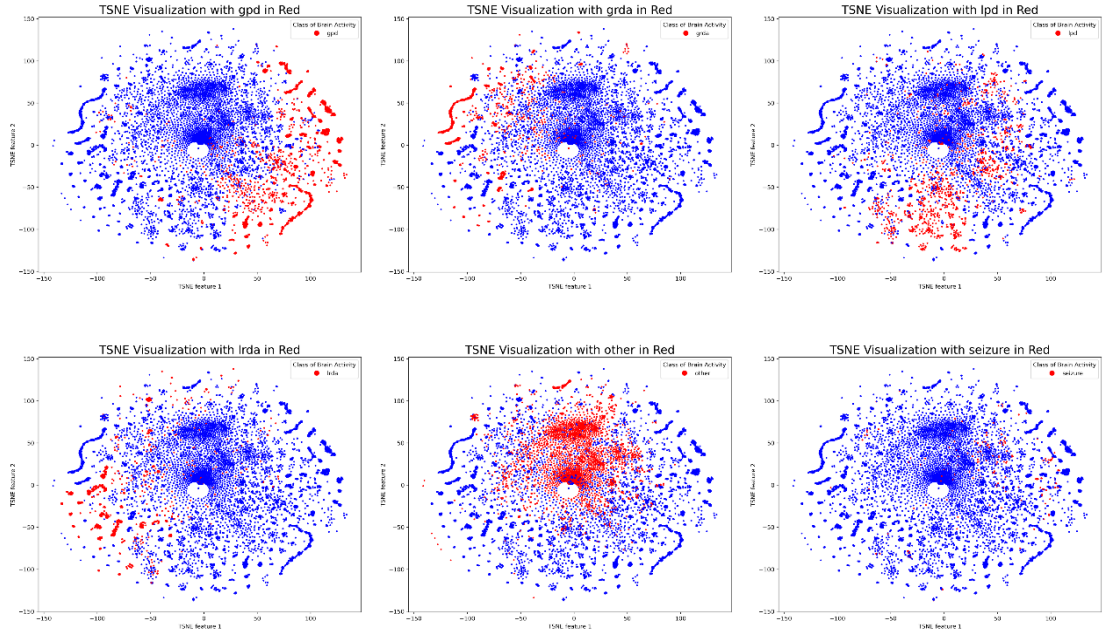

Figure 9 Single-class t-sne dimensional reduction visualization of EEG-SCFNet model

## 3.7   The training results on the CHB-MIT dataset

This experiment involves transferring the EEGNet and EEG-SCFNet models trained on the IIIC-Seizure dataset to the CHB-MIT dataset for transfer learning, thereby verifying the cross-dataset transfer learning performance of the models. The experiment uses the IIIC-Seizure dataset (16-channel 50s data) to train the model parameters and fine-tunes on the CHB-MIT dataset (16-channel 50s data). To compare the potential loss of transfer learning, the models are also trained from scratch on the CHB-MIT dataset. The comparison results are shown in Table 4. Figure 10 displays the confusion matrix of the prediction results from the 5-fold cross-validation of the four models, and Figure 11 shows the Macro ROC and Micro ROC situations of the four models.

The EEG-SCFNet model, after being trained from scratch on the CHB-MIT dataset, achieved a Balance Accuracy of 80.96%, and its average accuracy in 5-fold cross-validation reached as high as 91.11%. In comparison, the BIOT[8] model has a Balance Acc of 57%, and in the study by Taherisadr et al. [24], its average accuracy evaluated using 10-fold cross-validation was 89.63%. The traditional end-to-end model EEGNet, which was not modified, had a ROC AUC of only 50.07% when transferred from the IIIC-Seizures dataset to the CHB-MIT dataset, making the model completely unusable. However, the transferred EEG-SCFNet achieved a Micro TPR of 71.49%, Micro RUC of 79.08%, and Macro RUC of 83.60%. It can be seen that the SCFNet network architecture enables the feature extraction module to achieve better transferability, maintaining certain performance despite inconsistencies in channel leads, although there is a decline in performance.

Table 4 Comparison of model training results of CHB-MIT dataset

| Model | EEG_M | EEG_M_F | EEG | EEG_F |
|---|---|---|---|---|
| Acc | 80.96 | 71.49 | 78.13% | 49.97% |

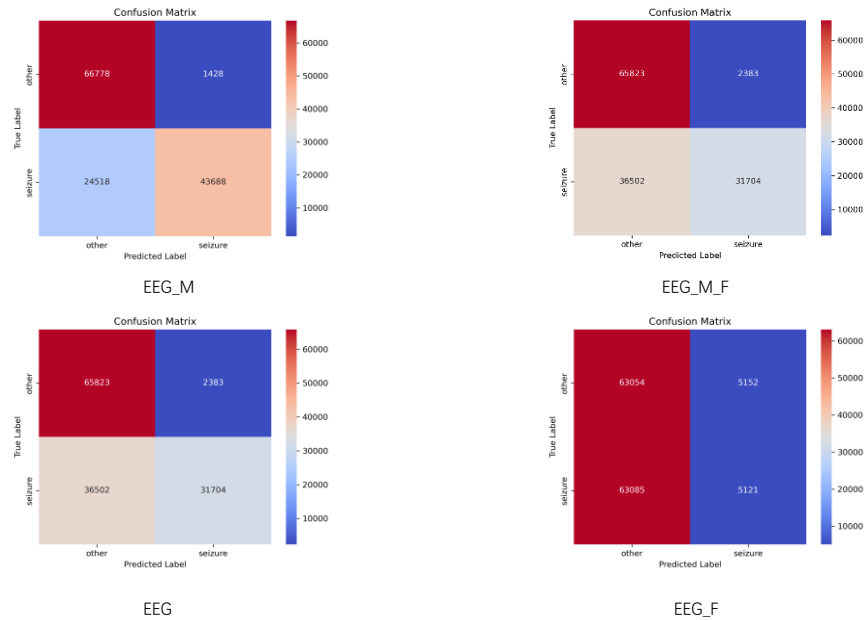

Figure 10 CHB-MIT dataset model training results comparison - confusion matrix

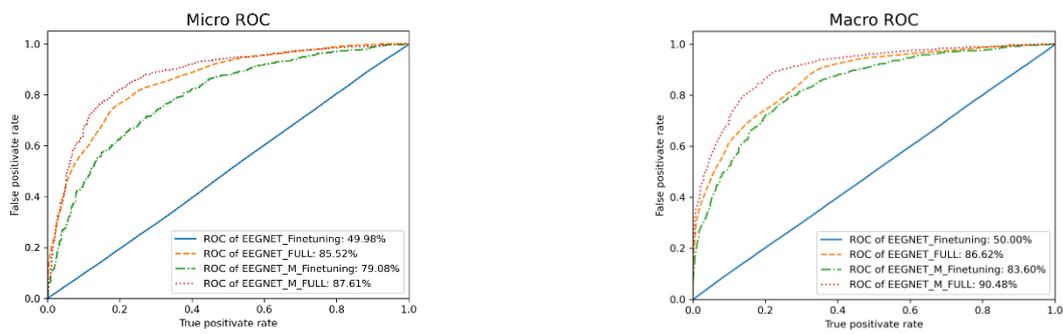

Figure 11 CHB-MIT dataset model training results comparison - ROC curve

# 4 Conclusion

Based on the single-channel classification network research in sleep stage classification, this paper proposes a single-channel feature extraction backend fusion network architecture that can effectively solve the problem of model transfer. Its effectiveness has been verified on the IIIC-Seizure and CHB-MIT datasets, and it has been found that this method can effectively improve training speed. On the IIIC-Seizure dataset, compared to the baseline model, it has improved by 4%, and compared to the improved model before, it has improved by 1.3%. In terms of cross-dataset transfer learning, using the IIIC-Seizure pre-trained model, by only training the classifier head, the accuracy on the CHB-MIT dataset reached 71.49%, which is a 9% drop from the model trained from scratch, while the original end-to-end method is completely unusable, greatly improving the model's transfer ability.

Firstly, this method only requires channel modification of the model's feature extraction layer and classifier. This idea should be applicable to other network structures, such as Transformer, CNN1D, or similar problems, such as on RGB-D datasets. It can also transform the original proven effective end-to-end models into network models that can transfer across channels. This paper only targets EEG detection and classification of harmful brain activities, and only experimented with RCNN as the basic network's modification effect. It did not conduct more experiments on other basic networks, nor did it expand experiments on other issues. In addition, since the feature extraction model only depends on a single channel and does not learn the mutual information between channels, a channel fusion module can be added before the classifier. This study attempted to add self-attention and CNN modules for channel fusion, but the effect was not good. The two directions mentioned above may become future research directions.